\documentclass[apj]{emulateapj}

\bibpunct {(} {)} {;} {a} {} {,}

\slugcomment{{\sc The Astrophysical Journal, accepted }  }

\begin{document}

\title{A Deep Photometric Look at Two of Andromeda's Dwarf Spheroidals: X and XVII}

\author{
Crystal Brasseur\altaffilmark{1},
Nicolas F.~Martin\altaffilmark{1},
Hans-Walter Rix\altaffilmark{1},
Mike Irwin\altaffilmark{2},
Annette Ferguson\altaffilmark{3},
Alan W. McConnachie  \altaffilmark{4}
Jelte de Jong  \altaffilmark{1}}

\affil{$^{1}$Max-Planck-Institut f\"{u}r Astronomie, 17 K\"{o}nigstuhl 69117 Heidelberg, Germany
\\$^{2}$Institute of Astronomy, Madingley Road, Cambridge, CB3 0HA, UK
\\$^{3}$Institute for Astronomy, University of Edinburgh, Blackford Hill, Edinburgh, EH9 3HJ, UK
\\$^{4}$Herzberg Institute of Astrophysics, National Research Council Canada, 5071 W.~Saanich Rd., Victoria, B.C., V9E~2E7, Canada
}

\begin{abstract}

We use deep wide-field photometry from the Large Binocular Camera to study the stellar and structural properties of the recently discovered Andromeda X and Andromeda XVII (And~X and And~XVII) dwarf galaxies. Using the mean apparent magnitude of the horizontal branch (HB), we derive distances of 621 $\pm$ 20 kpc to And~X and 734 $\pm$ 23 kpc to And~XVII, closer by $>$60 kpc than the previous estimates which were based on red giant branch (RGB) observations. Thus our results warrant against the use of the RGB tip method for determining distances to systems with sparsely populated RGBs, and show how crucial HB observations are in obtaining accurate distances in systems such as these. We find that And~X is a relatively faint (M$_V$ = $-$7.36), highly elongated ($\epsilon=0.48$) system at a distance of 174 $\pm$ 62 kpc from Andromeda. And~XVII is brighter (M$_V$ = $-$8.61) with an M31-centric distance of 73 kpc which makes it one of the closest satellites to Andromeda. Both galaxies are metal-poor: we derive $\langle[Fe/H]\rangle$=$-$2.2 for And~X, while And~XVII shows $\langle[Fe/H]\rangle = -2.0$, consistent with the relation of higher luminosity dwarfs being more metal-rich. Additionally, both galaxies show considerable intrinsic spreads in metallicity (0.2 and 0.3 dex for And~X and And~XVII respectively), consistent with multiple stellar populations.

\end{abstract}

\section{Introduction}
\label{sec:intro}

Our view of the low-mass end of galaxies has drastically broadened over the last decade with the discovery of systems now a hundred times fainter than previously known. Through its proximity, the Local Group affords a resolved look at the individual stars of dwarf galaxies, playing a crucial role in understanding their star formation histories.  Initially, Local Group observations were at odds with the standard cosmological framework of Cold Dark Matter, which predicts hundreds of halos to surround massive galaxies such as the Milky Way and Andromeda, however only a few tens of dwarf galaxies are observed (e.g., \citet{Klypin1999}; \citet{Moore1999}). Now, evidence has been mounting that star formation in low-mass halos becomes inefficient, eventually dropping below a limiting threshold, thus potentially explaining this `missing satellite' problem (e.g. \citet{Maccio2010}; \citet{Ki2010}; \citet{Coop2010}). 

Even with this limiting threshold, many open questions remain: what is the minimum stellar mass we can associate with a dark matter halo? Are all known dwarf galaxies bound systems? Are these faint galaxies all multiple stellar populations and self-enriched? We are limited by the paucity of examples with sufficient structural and population parameters. Therefore, looking deeply at satellites of the two nearest galaxies, the Milky Way and Andromeda (M31), is important for overcoming selection effects and investigating environmental differences. 

In this context we present here deep photometry of Andromeda satellites which we use to obtain basic parameters of two recently discovered dwarf galaxies, Andromeda~X and Andromeda~XVII (And~X and And~XVII). And~X was initially found in a dedicated scan of M31 in the Sloan Digital Sky Survey by \cite{Zucker2007}, while And~XVII was found in observations of the inner $\sim$50 kpc of Andromeda, performed with the Wide Field Camera on the Isaac Newton Telescope (\citealt{Irwin2008}).  From observations of the upper magnitudes of the RGB, \cite{Zucker2007} and \cite{Irwin2008} derived luminosities, metallicities and distances to each galaxy based on the RGB. We argue that, with their sparsely sampled RGBs, the absence of sub-horizontal branch (HB) photometry for these systems is detrimental to the accuracy of their distance estimates.

We have used the wide field imager on the Large Binocular Telescope (LBT) to obtain photometric observations that are deep enough to reach the HB. These observations increase the accuracy of the previously derived distances by using the mean HB magnitude, and the derived structural parameters through the increased star counts. In the following sections we briefly describe the data reduction and present the calibrated colour-magnitude diagrams (CMDs) in \textsection \ref{sec:observations}. The structural parameters are derived in \textsection \ref{sec:structuralparameters}.   Our methods for deriving the distances, metallicities and luminosities are described in \textsection \ref{sec:distances}, \textsection \ref{sec:metallicities}, and \textsection \ref{sec:luminosities}, respectively. Finally, a short summary of our results is presented in \textsection \ref{sec:summary}.

\begin{figure*}
\includegraphics[width=16cm]{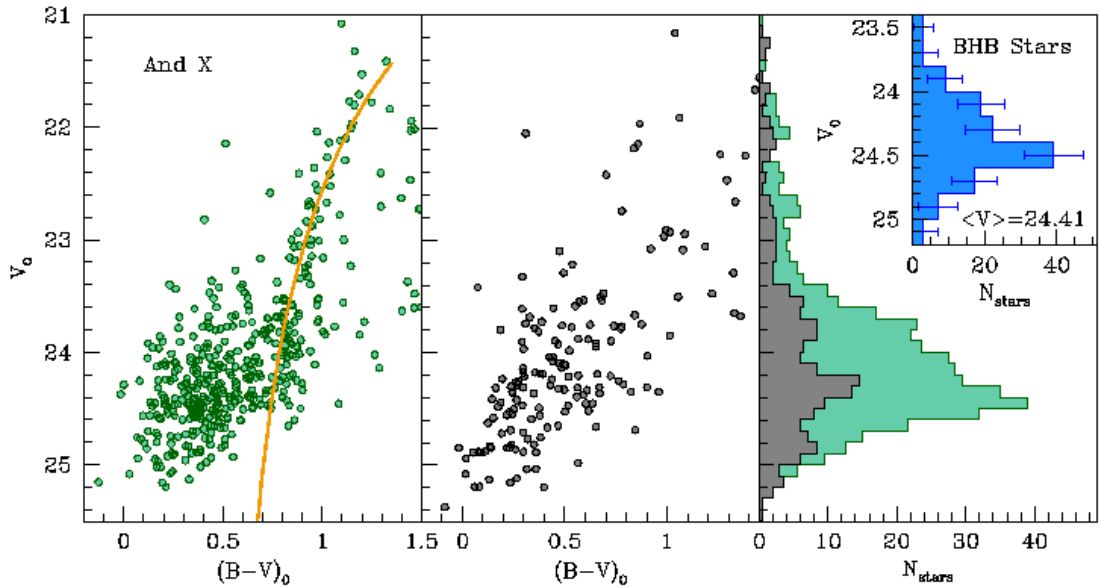}
\caption{  \textit{Left panel:} Extinction-corrected CMD of stars within an elliptical region ($a=1.5^{\prime}$, $e=0.48$, and P.A.= 46$^{\circ }$ ) centered on And~X. Overlaid is an isochrone corresponding to [Fe/H]=$-$2.2, [$\alpha$/Fe]=+0.4, which has been shifted to the distance of And~X determined in \textsection \ref{sec:distances}. \textit{Center panel:} The underlying CMD contamination in a region of equal coverage to that of the left panel, but centered on a point in the same field offset from the dwarf. \textit{Right panel:} The V-band LF of the region centered on And~X is represented by the larger turquoise histogram, while the inner gray histogram represents a background region. In the upper right corner, the inset blue histogram shows the mean background corrected LF of blue HB stars based on 1000 realizations (see \textsection \ref{sec:distances} for details). Our derived background and completeness-corrected mean magnitude of the HB is, $<V_{HB}>$=24.41, and results in an absolute distance modulus of $(m-M)_0$ = 24.01 $\pm$ 0.07. }
\label{andx}
\end{figure*}

\section{Observations}
\label{sec:observations}

In 2008, we received Large Binocular Camera (LBC) observing time on the LBT to observe the two dwarf galaxies, And~X and And~XVII. Our observations with the blue eye of the telescope aimed to reach a significant signal-to-noise ratio (S/N) at the level of the HB of each galaxy. To accomplish this, observations of And~X comprise 10 $\times$ 2 min and 8 $\times$ 2 min exposures in the $V$ and $B$ bands respectively and observations of And~XVII comprise 10 $\times$ 4 min in the $V$ band and 9 $\times$ 2 min exposures in the $B$ band with a seeing of $\approx$1 arcsec.   By taking a series of exposures, each star was detected multiple times, thereby helping to improve the precision of the final photometry.  Additionally, several Landolt standard star fields were observed in order to calibrate the final photometry.  

Once observed, all images were pre-processed with MPIAPhot, a set of MIDAS routines developed to
reduce wide-field imager data\footnote{http://www.mpia.de/HIROCS/mpiaphot.pdf}.
Bias images and twilight flats
observed during the same nights as the And~X and XVII observations were used to generate master bias and flat field images. After de-biasing
and flat-fielding the frames, they were divided into individual images for
their four chips. Both dwarf galaxies are surrounded by bright stars, producing a
varying background level from chip to chip. We therefore focus only on chip 2 which
contains the dwarf galaxy in both cases. Astrometry on single science frames was
performed by comparison with the UCAC catalogue (\citet{Z00}).

\begin{figure*}
\includegraphics[width=16cm]{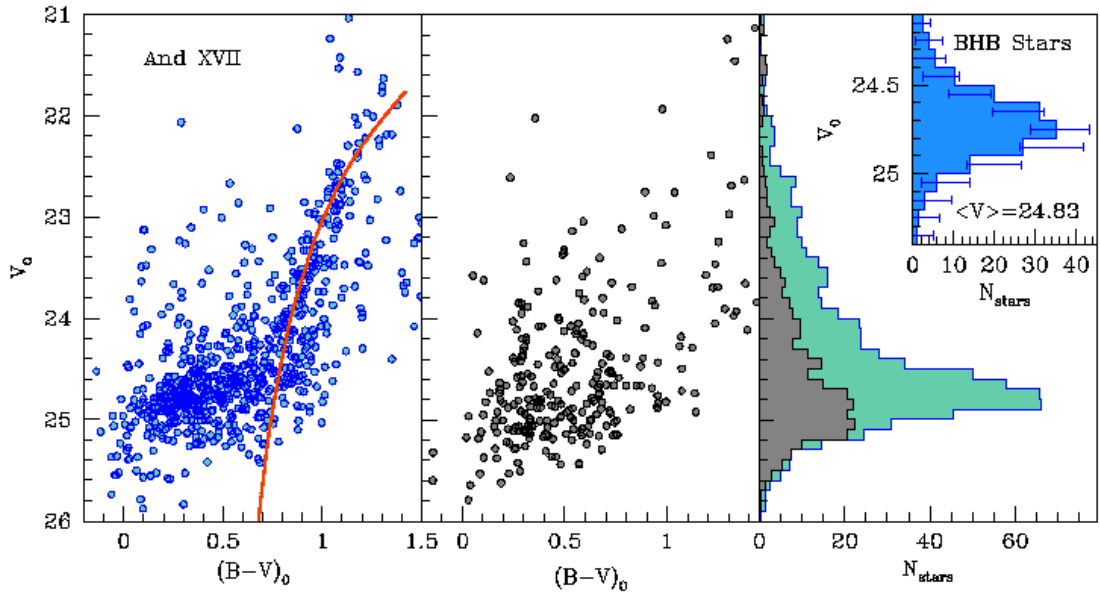}
\caption{ \textit{Left panel:} Extinction-corrected CMDs of stars within an elliptical region ($a=1.5^{\prime}$, $e=0.36$, and P.A.= 127$^{\circ }$ ) centered on And~XVII.  Overlaid is an isochrone corresponding to [Fe/H]=$-$2.0, [$\alpha$/Fe]=+0.4 which has been shifted to the distance of And~XVII determined in \textsection \ref{sec:distances}. \textit{Center panel:} The underlying CMD contamination in a region of equal coverage to that of the left panel, but centered on a point in the same field offset from the dwarf. \textit{Right panel:} The V-band LF of the region centered on And~XVII is represented by the larger turquoise histogram, while the inner gray histogram represents a background region. In the upper right corner, the inset blue histogram shows the mean background corrected LF of blue HB stars based on 1000 realizations (see \textsection \ref{sec:distances} for details). Our derived mean magnitude of the HB is  $<V_{HB}>$=24.83, and results in a absolute distance modulus of 24.36 $\pm$ 0.06. }
\label{andxvii}
\end{figure*}

Instrumental magnitudes for all stars were obtained using the point spread function (PSF) modeling and fitting techniques in the DAOPHOT/ALLSTAR/ALLFRAME packages (\citet{Stetson1987}; \citet{StetsonHarris1988}).  In essence, these programs work by detecting stars on a specific image, building a model PSF from a few isolated, bright stars and then subtracting this PSF from all stars detected. In the final catalog of detections, DAOPHOT provides a sharp index which measures the degree to which an object's intrinsic angular radius differs from that of the model PSF, i.e., real stars have a sharp value near zero. Detections with large positive sharp values have larger characteristic radii compared to the PSF model and are most likely resolved galaxies where as detections with very negative sharp values have apparent radii smaller than the seeing profile, and are likely to be cosmic rays. We apply a cut of $\mid$sharp$\mid$ $<$ 1 and $\mid$chi$\mid$ $<$ 2 to extract stars from the final photometry catalog for our analysis, which have a high probability of being real stars.

To obtain the photometric equations, the Landolt standard fields were used to solve for the zero points and colour terms which were then applied to the science fields.

We de-reddened our photometry by applying a uniform correction according to the \cite{Schlegel98} maps, assuming an extinction to reddening ratio $A_V / E(B-V) = 3.1$ and $A_B / E(B-V) = 4.1$.

\section{Properties of the Satellites from Colour-Magnitude Diagram Maps}
\label{sec:properties}

We present our resulting CMDs in Figures \ref{andx} and \ref{andxvii}, plotting all stars within two half-light radii of each dwarf (as determined in \textsection \ref{sec:structuralparameters}). 
Our observations of each galaxy extend from the tip of the RGB, down to the HB to a $S/N$ of 5 at $V \approx$25; for And~XVII, this is a dramatic improvement over the previous CMD from the discovery paper which contained only the upper part of the RGB (\citealt{Irwin2008}). Both galaxies show well defined HBs, and interestingly,  And~XVII shows a significant population of blue HB stars, $(B-V)<0.3$, similar to those of metal-poor Galactic globular clusters.  And~X and And~XVII both show steep RGBs and no evidence of a red clump, which indicates that the stellar populations are relatively metal-poor. Without observing the main-sequence turnoff it is not possible to derive a precise age of any stellar system. However, it is possible to constrain whether a system is young or old based on other features in the CMD. For example, no population of thermally pulsating asymptotic giant branch stars is seen in our CMDs, therefore we can rule out the presence of a very young stellar population. 

In the central panel of Figures \ref{andx} and \ref{andxvii}, we show the underlying CMD contamination in a region of equal area more than 7 arcmin away from the dwarf center, on the same chip. This gives a qualitative view of the contamination from the background, which is made up of Milky Way foreground, a few M31 halo stars and unresolved background galaxies masquerading as stars around the same magnitude as the HB. Due to the problematic position of these contaminates in the CMD, the effect of these galaxies must be carefully removed when determining distances. It is on the basis of these CMDs that we derive greatly improved estimates of their distance and structure.

\begin{figure}
\includegraphics[width=8cm, angle=270]{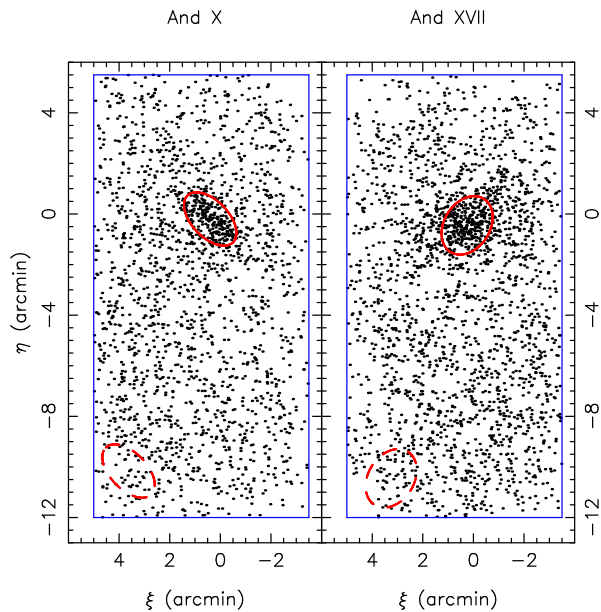} 
\vspace{1cm}
\caption{ Spatial distribution of stars with colours consistent with being RGB and HB members around And~X (left panel) and And~XVII (right panel) in chip 2 of the LBC. Both galaxies appear as obvious stellar over-densities. The solid ellipses correspond to the region within one half-light radii of each dwarf galaxy, assuming the structural parameters listed in Table 1. The dashed ellipses correspond to the areas we have chosen for background correction.
}
\label{andiesmaps}
\end{figure}

%%%%%%%%%%%%%%%%%%%%%%%%%%%%%%%%%%% Tables
%%%%%%%%%%%%%%%%%%%%%%%%%%%%%%%%%%%

\begin{deluxetable*}{l c c}
\tabletypesize{\footnotesize}
\tablecaption{Properties of And~X and And~XVII. \label{cluster_params}}
\tablewidth{0pt}
\tablehead{ \colhead{Parameter} & \colhead{And~X} & \colhead{And~XVII} }
\startdata
$\alpha$ (J2000)          & 1:06:35.3 ($\pm$10")  & 0:37:08.0 ($\pm$10")   \\
$\delta$ (J2000)          & +44:48:03.8  ($\pm$10")       & +44:18:53.0 ($\pm$10")\\
PA (North to East)        & 46\,$^{\circ}$ $\pm$ 5    & 122\,$^{\circ}$ $\pm$ 7\\
$\epsilon=1-b/a$                   & 0.44 $\pm$ 0.06     & 0.27 $\pm$ 0.06\\
$r_h (arcmin)$            &1.30 $\pm$ 0.10       & 1.24 $\pm$ 0.08\\
$E(B-V)$\tablenotemark{1} & 0.128               &  0.0728 \\

$(m-M)_0$                 & 24.01 $\pm$ 0.07    &  24.36 $\pm$ 0.06 \\
D (kpc)                   & 621 $\pm$ 20      &  734 $\pm$ 23 \\
$r_{M31}$ (kpc)\tablenotemark{2} & 174 $\pm$ 29      &  73 (min 49, max 100) \\
$M_V$                     &$-$7.36 $\pm$ 0.07 (photometric) $+/$ 0.92 (systematic) &  $-$8.61$\pm$ 0.07 (photometric) $+/$ 0.30 (systematic)\\
$\langle[Fe/H]\rangle_{13 Gyr}$\tablenotemark{3}& $-2.2$          &  $-2.0$\\
$\langle[Fe/H]\rangle_{10 Gyr}$\tablenotemark{3}& $-2.1$          &  $-1.9$\\
$\langle[Fe/H]\rangle_{[\alpha/Fe]=0}$\tablenotemark{4}& $-2.1$          &  $-1.7$\\
Intrinsic [Fe/H] spread & 0.23$\pm$0.04 dex         & 0.31 $\pm$0.03 dex\\

\enddata
\tablenotetext{1}{\citet{Schlegel98}}
\tablenotetext{2}{We assume a distance to M31 of 783$\pm$25 kpc (\citet{M2005}).}
\tablenotetext{3}{Here we have adopted [$\alpha$/Fe]=+0.4.}
\tablenotetext{4}{Here we have adopted an age of 13 Gyr.}
\end{deluxetable*}

\begin{figure}
\includegraphics[width=7cm, angle=270]{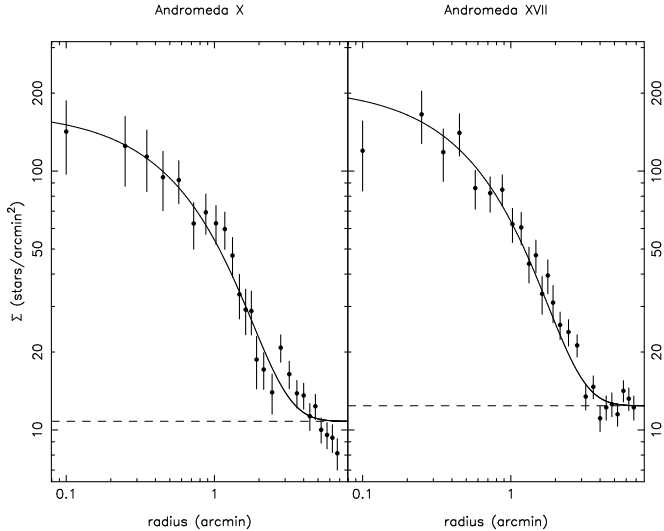}
\vspace{1cm}
\caption{
Stellar profiles of And X and And XVII. In both panels, the stellar
density measured in fixed elliptical annuli from the LBC data is shown
as dots, using the best ellipticity, position angle and centroid found by
the fitting algorithm (the uncertainties are derived assuming Poisson
statistics). These data points are corrected for incompleteness produced
by CCD gaps and the halo of bright stars. The full lines represent the
best exponential radial density models constructed from the best fit half-light radii and background
levels, which are represented by the thin dashed lines. Although they are not
fits to the data points, the profiles are in good agreement with them.
}
\label{andiesmaps2}
\end{figure}

\subsection{Structural Parameters}
\label{sec:structuralparameters}

The structural parameters of the two dwarf galaxies were determined by following the
technique presented in \citet{MdJR2008}, and later updated in \citet{Martin2009} to 
account for the halos of bright stars and the
limited spatial extent of the data. We refer the reader to these papers for more details, and, in particular, section 2.1 of \citet{MdJR2008} for a detailed description of the model used. To
summarize, stars with a location in the color-magnitude space that is compatible
with either the RGB, or the HB of the system (shown in Figure \ref{andiesmaps}) are used to constrain the best exponential
radial profile for each dwarf galaxy which is shown in Figure \ref{andiesmaps2}. The fit relies on the maximum likelihood
technique to yield the best estimate of the centroid $(\alpha_0,\delta_0)$, the
half-light radius ($r_h$), the ellipticity\footnote{The ellipticity of the
system is here defined as $\epsilon = 1-b/a$, with $b$ the scale-length of the system
along its minor axis and $a$ that along its major axis.} of the system ($\epsilon$),
the position angle (P.A.) of the major axis of the system from north to east, as
well as the contamination of foreground/background sources, taken as a constant over
the field of view. The best values of these parameters and their corresponding
uncertainties (determined as explained in Martin et al. 2009) are listed in Table~1.

Investigating the spatial distribution of different stellar sub-populations, we found no significant difference between the metal-poor and metal-rich stars in either galaxy. 
For both galaxies, a Kolomogorov-Smirnov test implies an inconclusive 25$\%$ probability that both metal-rich and metal-poor samples have the same radial distributions.

\begin{figure*}
\includegraphics[width=8cm]{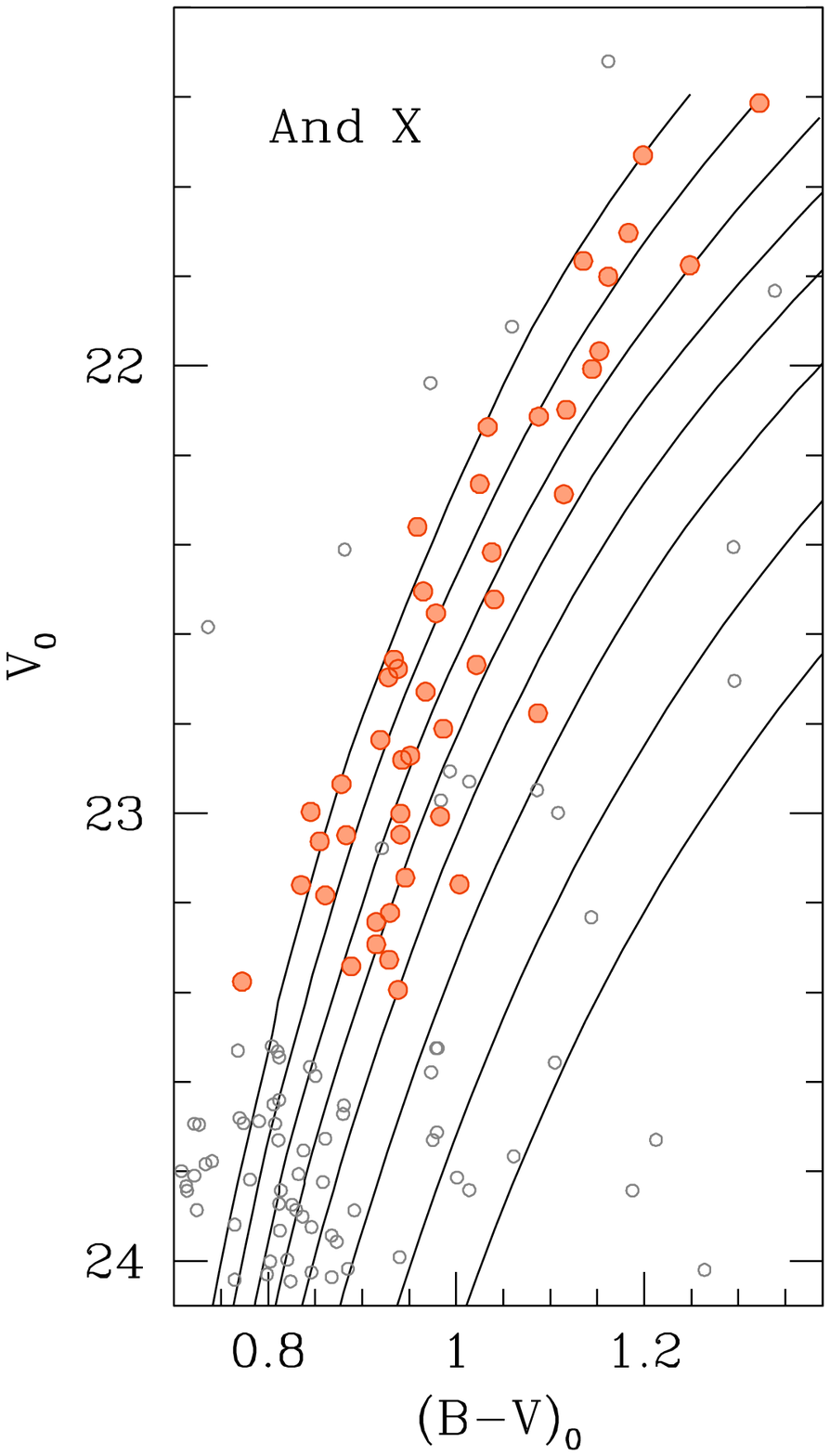}
\includegraphics[width=8cm]{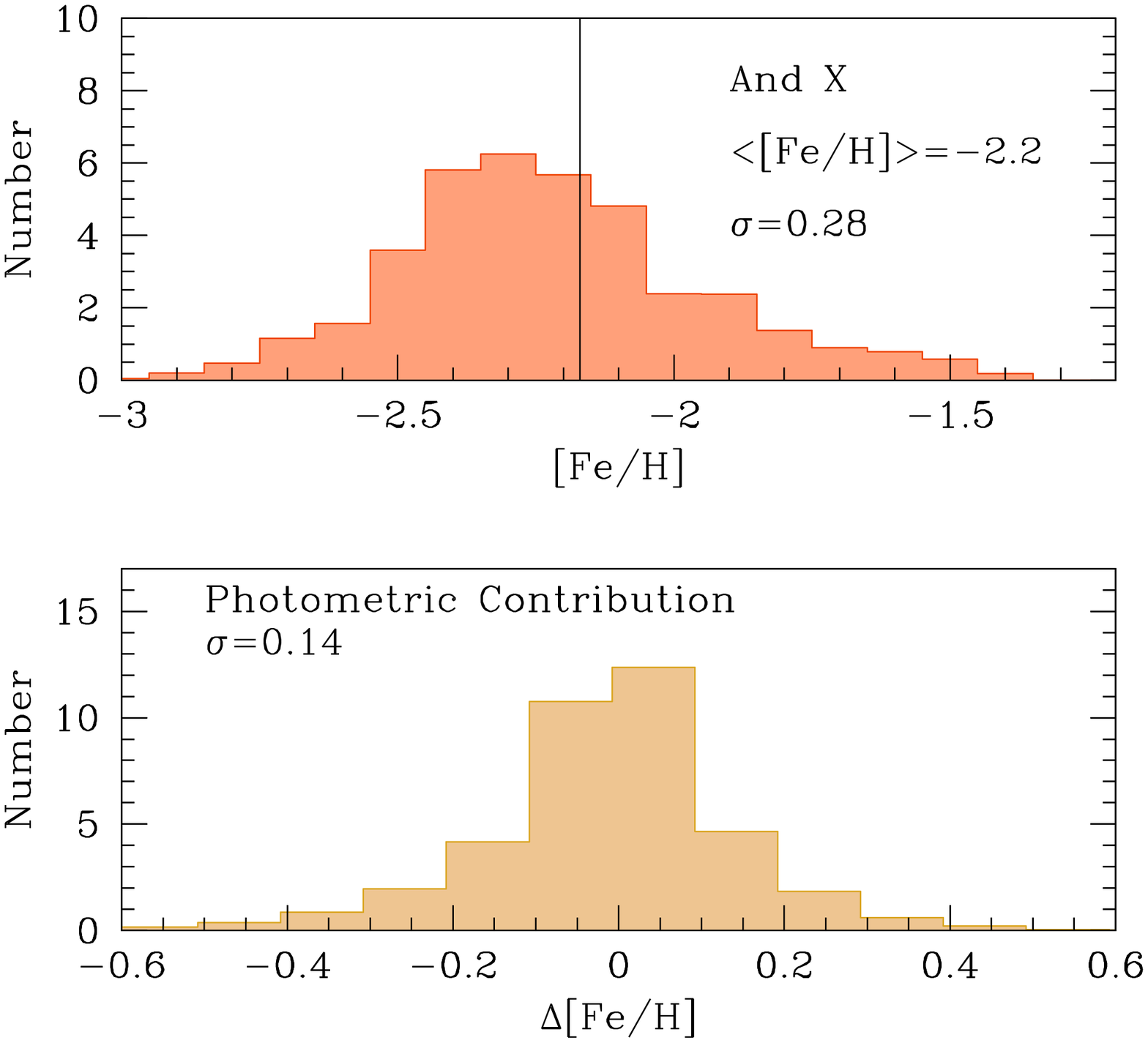}
\caption{ \textit{Left:} The red giant branch stars of And~X overlying 13 Gyr old \citet{Dotter} isochrones. All stars of the CMD are shown in open gray circles, and those selected to be RGB members are represented by filled pink circles. From left to right, the isochrones range in metallicity from $-2.5$ to $-1.1$ and are shown in increments of 0.2 dex. \textit{Upper Right:} The metallicity distribution of the RGB stars shown in the left panel, averaged over 1000 re-samplings based on their photometric errors (see \textsection \ref{sec:metallicities} for details). \textit{Lower Right:} The distribution of the individual $\Delta$[Fe/H]s, averaged over 1000 RGB re-samplings. The spread of this distribution is the expected [Fe/H] width of the RGB due to photometric errors alone. Subtracting this off the observed spread in quadrature, we arrive at a intrinsic metallicity spread of 0.23 $\pm$ 0.04 dex. }
\label{mdandx}
\end{figure*}

\subsection{Distances}
\label{sec:distances}

Previous distance determinations to both And~X and And~XVII (\citealt{Zucker2007}; \citealt{Irwin2008}) were made using the $I$-band RGB tip magnitude which marks the core helium flash of low-mass Population II stars. The advantages of using the tip as a distance indicator have long been recognized: both empirical and theoretical studies find the RGB tip I-band magnitude to be virtually constant for systems older than $\sim$ 2 Gyr with metallicities between $-2.2<{\rm [Fe/H]}<-0.7$. However, this method is only reliable when the RGB tip is well populated in the photometry. For systems with sparsely populated RGBs, the location of the tip becomes increasingly difficult to determine, as is the case for And~X and And~XVII. Thus, when applying the RGB tip method to fainter galaxies such as these, critical stars around the tip of the RGB may not be present leading to an overestimated distance. By obtaining photometry which reaches the HB, we circumvent these uncertainties, leading to a more confident distance modulus. 

The absolute mean magnitude of the HB is well established through RR Lyrae measurements to produce the following calibration of \citet{CacciariClementini}:

$M_V(HB) =  (0.23\pm 0.04)([Fe/H]+1.5)+(0.59\pm0.03).$

The mean metallicity of each galaxy determined in \textsection \ref{sec:metallicities}, implies $M_{V}$(HB) = 0.43 $\pm$ 0.06 for And~X and $M_{V}$(HB) = 0.47 $\pm$ 0.05 for And~XVII.

The distance modulus then results from subtracting these values from the mean $apparent$ magnitude of the HB, $\langle V_{HB}\rangle$. To quantify $\langle V_{HB}\rangle$, and its associated error from our data, including the individual photometric uncertainties, we use a Monte Carlo approach. 
We assume a bivariate Gaussian distribution for the photometric errors of each star, with $\mu_V$ and $\mu_{B-V}$ as the position of each star and the standard deviation set as the photometric error, $\sigma_V$ and $\sigma_{B-V}$, associated with each individual star.  To test whether our photometric uncertainties are truly Gaussian, we compared the photometry presented in the paper with that obtained by applying a completely different algorithm, namely that presented by \cite{Irwin2001}. The two algorithms are completely independent (PSF versus aperture photometry) yet the distribution of differences in magnitudes measured by the two methods, normalized by the sum of the photometric uncertainties added in quadrature yield a Gaussian distribution. We are therefore confident that our photometric uncertainties are indeed Gaussian for the range of magnitudes 21-24.5.

\begin{figure*}
\includegraphics[width=8cm]{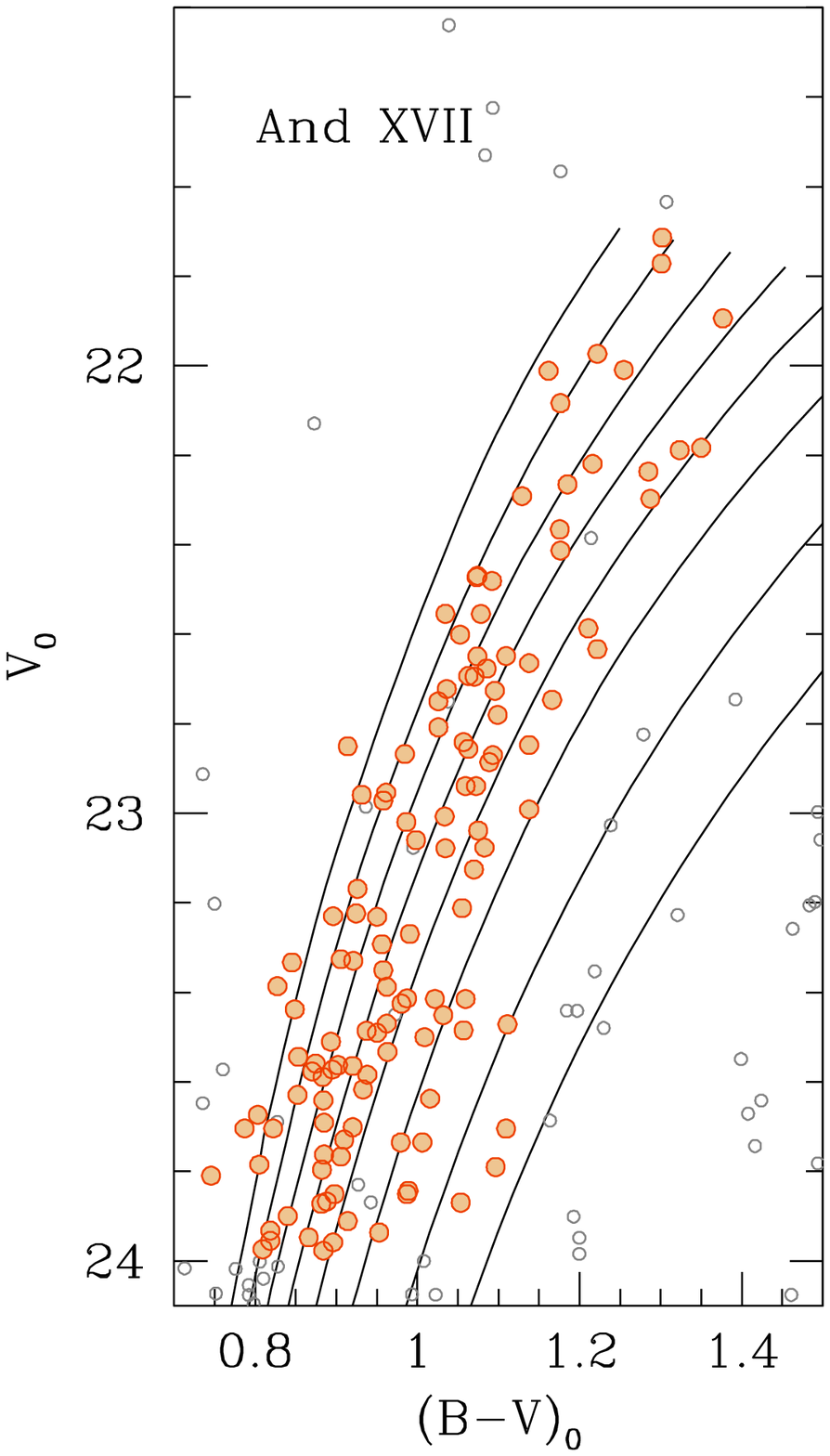}
\includegraphics[width=8cm]{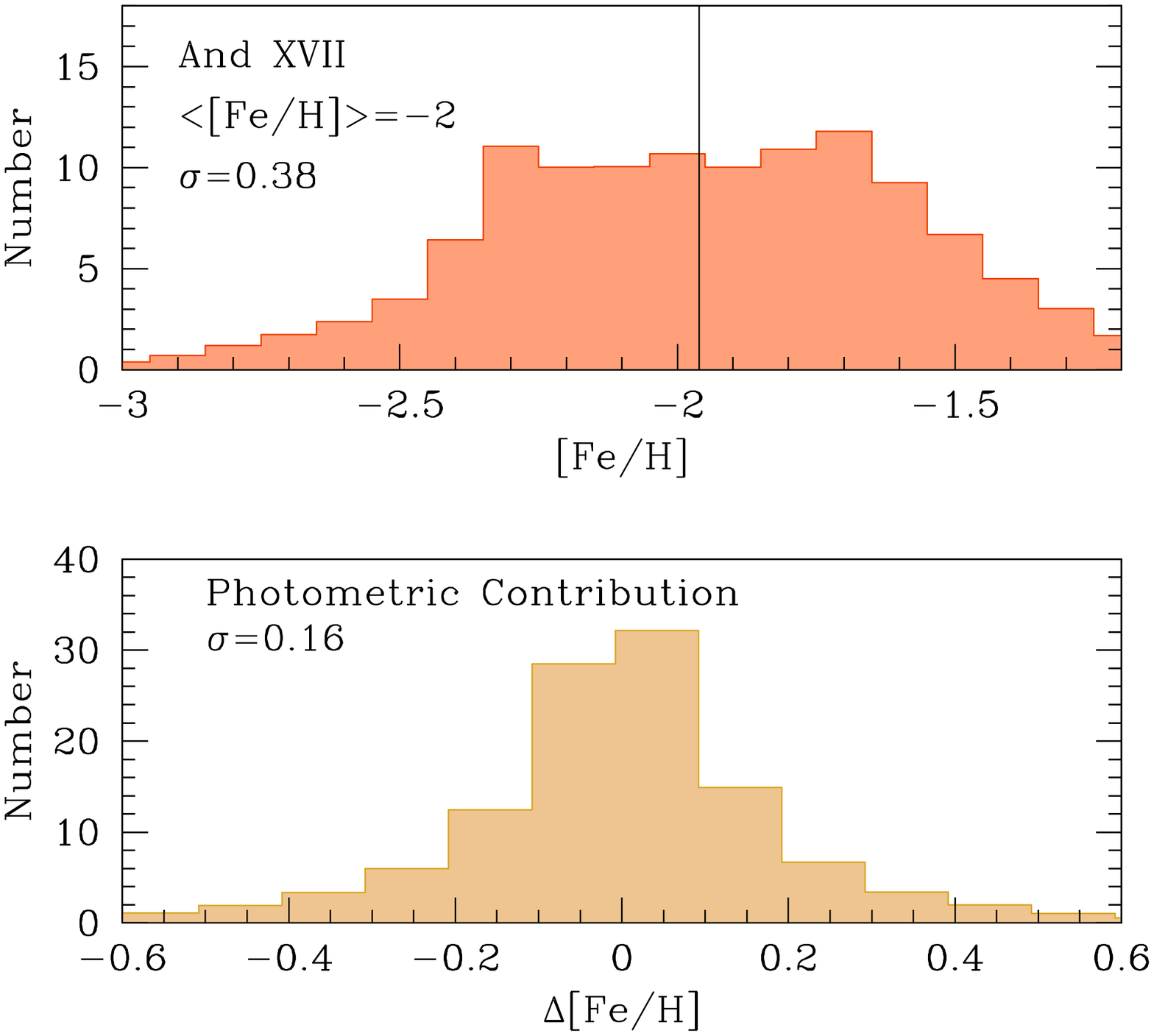}
\caption{Same as for Figure \ref{mdandx}, except for And~XVII. The measured intrinsic metallicity spread is here 0.31 $\pm$ 0.03 dex.}
\label{mdandxvii}
\end{figure*}

We generate 1000 HBs by drawing points at random from the Gaussian photometric error distribution of each star in the CMD, and applying a colour cut of $(B-V)<$ 0.5 to produce 1000 HB realizations. For each realization, we generate a background corrected $V$-band HB luminosity function (LF). 

To obtain the apparent magnitude of the HB we must account for the fact that the faint end of the HB is being affected by completeness, which could skew the mean to brighter magnitudes. We therefore compute the completeness function of our observations using the globular cluster, M92, and then forward model this effect on a Gaussian fit to the blue HB histogram (upper right panel of Figures 1 and 2, thus finding the mean magnitude of a deformed Gaussian which was the product of a Gaussian and our completeness function. This method results in 
the mean apparent magnitude of the HB being 0.03 mag fainter compared to the value obtained when not taking the completeness of the data into account. It is therefore only a mild correction.

In the right panels of Figures \ref{andx} and \ref{andxvii}, we show the mean background corrected HB LF of our 1000 realizations.  From the distribution of the means of the 1000 background and completeness-corrected HB LFs, we compute $\langle V_{HB}\rangle$=24.41$\pm$ 0.03 for And~X and $\langle V_{HB}\rangle$=24.80$\pm$ 0.03, where the standard deviation of these distributions are taken as the uncertainty on $\langle V_{HB}\rangle$.

Combining our determinations of absolute and apparent magnitude, we derive distance moduli of $(m-M)_0$=23.98 $\pm$ 0.07 to And~X and 24.33 $\pm$ 0.07 for And~XVII. These correspond to line-of-sight distances of 621 $\pm$ 20 kpc and 734 $\pm$ 23 kpc for And~X and And~XVII, respectively.  To determine M31-centric distances, we adopt the distance modulus of M31 to be $(m-M)_0$ = 24.47 $\pm$ 0.07, following \citet{M2005}, or 783 $\pm$ 25 kpc. With the angular separation of And~X from the center of M31 being 5.5\,$^{\circ}$, this gives a projected separation of 75 kpc at the distance to M31.  Combining this with the line-of-sight distance derived from the HB, we obtain an M31-centric distance of $r_{M31}$ = 168 $\pm$ 29 kpc for And~X, well within M31's estimated virial radius of 300 kpc (\citealt{K2002}). And~XVII lies at a projected distance of 3$^{\circ}$.2 from the center of M31, making it, in projection, one of the innermost known members of the satellite system. This translates to a projected separation of 44 kpc at the distance of Andromeda, and a distance of $r_{M31}$ = 73 from the center of M31. Due to the non-Gaussian shape of the probability distribution function of the M31-centric distance, we do not quote errors in the same way as for And X. We arrive at a minimum distance of 44 kpc (the projected separation) and a maximum distance of 100 kpc.

\subsection{Metallicities}
\label{sec:metallicities}

The apparent width of the giant branch in the CMDs of And~X and And~XVII seen in Figures \ref{andx} and \ref{andxvii}, naively suggests that the galaxies are composed of stars with a range of metallicities.  However, we must carefully explore which fraction of the observed width is due to photometric errors and which is due to intrinsic metallicity variations.

We apply a colour cut to extract RGB stars from each dwarf, and a faint limit such that we minimize the contamination from HB stars. Our background corrected selection is shown by the orange filled points in Figures \ref{mdandx} and \ref{mdandxvii}. Although it is correct that the background is statistically
removed, it is still necessary to apply the initial colour cut: since the metal poor isochrones become increasingly closer together, we found that only a few background blue stars which escaped removal were able to significantly skew the mean metallicity.

The metallicity estimates of individual RGB stars were obtained by interpolating (or extrapolating when necessary) their position in colour-magnitude space in a grid of \citet{Dotter} isochrones at the distance moduli determined above. A representation of how this is done is shown in the left panels of Figures \ref{mdandx} and \ref{mdandxvii}, where we have overlaid the RGB stars onto isochrones of different metallicity to show the relative positions in the ($B-V$)$_0$-$V_0$ plane. To determine the metallicities we have used 13 Gyr, $[\alpha/Fe]$ = +0.4 isochrones ranging from [Fe/H] = $-$2.5 to [Fe/H] = +1.0, in steps of 0.1 dex. We then correct for the background by removing the nearest RGB neighbour to each background RGB star from the computation, although with and without correction produces an identical resulting mean. In this way both galaxies are determined to be metal-poor, with derived mean metallicities of $\langle$ [Fe/H]$\rangle$ = $-$2.2 for And~X and $\langle$ [Fe/H]$\rangle$ = $-$2.0 for And~XVII. The corresponding isochrones to these mean metallicities are overlaid on the CMDs in Figures \ref{andx} and \ref{andxvii}, and show good agreement.

Since adjustments to the distance moduli used in the above analysis will affect these metallicity determinations, we re-run the analysis for 1$\sigma$ shifts our determined $(m-M)_0$. These shifts in distance translate to $\approx$0.05 dex differences in metallicity (more metal-poor for larger distance moduli and more metal-rich for smaller distance moduli).

Without observing down to the main-sequence turnoff, the precise ages of And~X and And~XVII are unknown.
We have made the assumption that the stellar populations in both galaxies are predominately old, based on evidence of other dSphs whose star formation histories have been measured (e.g., \citet{Mateo1998}; \citet{Harbeck2001}) and the lack of thermally pulsating asymptotic giant branch stars in the CMD which would indicate the presence of a young population of stars. Even if our age assumption is inaccurate, for a shift in age from 13 Gyr to 10 Gyr, we derive the mean metallicities for our galaxies to be only 0.1 dex more metal-rich. Similarly to the uncertainty in age, there are also currently no direct constraints on the $\alpha$-enhancement of And~X or And~XVII.
In Milky Way studies, high resolution stellar spectra have shown dSphs of low metallicity (Draco, Ursa Minor, and Sextans with [Fe/H]= $-$2.00 $\pm$ 0.21, $-$1.90 $\pm$ 0.11, and $-$2.07 $\pm$ 0.21 respectively) to be slightly enhanced in $\alpha$-elements, 0.02 $\lesssim$ [$\alpha$/Fe] $\lesssim$ 0.13
(\citet{Shetrone2001}).  Since there are currently no $\alpha$-abundance measurements for $any$ Andromeda dSphs, we run the analysis for [$\alpha$/Fe]=0, in addition to the [$\alpha$/Fe]=+0.4. The solar alpha abundance yields more metal-rich estimates of $\langle$ [Fe/H]$\rangle$=$-$2.1 for And~X and $\langle$ [Fe/H]$\rangle$=$-$1.7 for And~XVII. Therefore we conclude that reasonable shifts in distance, age and alpha-abundance do not drastically affect our metallicity scale.

In order to determine the mean metallicity and the error-corrected metallicity spread, we use a Monte Carlo approach. We generate 1000 RGBs by drawing points at random from the Gaussian error distributions for each star in the CMD, and applying our RGB colour cut. For each of the 1000 realizations, we remove background stars as before and interpolate each star's colour-magnitude position through the grid of isochrones.  The resulting metallicity distribution of each dwarf, averaged over the 1000 realizations, is shown in the upper right panels of Figures \ref{mdandx} and \ref{mdandxvii}. To compute the error corrected spread of this distribution, we interpret the photometric errors as metallicity differences ($\Delta$[Fe/H]s). 
In the lower right panels of Figures \ref{mdandx} and \ref{mdandxvii} we show the distribution of the $\Delta$[Fe/H]s averaged over the 1000 realizations. The spread of this distribution is then the contribution of photometric errors to the observed metallicity spread. Thus, we subtract in quadrature the expected width from the observed width, to derive the intrinsic metallicity spreads of 0.23 $\pm$ 0.04 dex for And~X and 0.31 $\pm$ 0.03 dex for And~XVII.

It should be noted that the determination of the metallicity spread is more robust than the absolute values as isochrones based on different stellar evolutionary models will not yield the same mean metallicities as derived here. Also due to the small color differences between metal-poor isochrones, the resulting spread in the derived metallicity will change.  The important result to take away from this section is that these systems are both metal-poor ($\langle$ [Fe/H]$\rangle < -$1.9) and show significant metallicity spreads, consistent with multiple stellar populations.

\subsection{Luminosities}
\label{sec:luminosities}
To determine the total luminosity of each dwarf we again use a Monte Carlo approach. We generate 1000 realizations of the CMD by drawing points from the Gaussian error distributions of each star and then apply the same RGB cut as in the metallicity section. For each realization, we calculate the mean $L_{RGB}$ by summing the flux contribution of each RGB star, and correcting for the expected contribution of the background. This giant branch luminosity then needs to be augmented by the contribution of stars too faint to be observed. In order to do this we use the \citet{Dotter} 13 Gyr, [$\alpha$/Fe]=+0.4 LFs of the mean metallicity of the dwarf to determine the fraction of missing light (i.e., (RGB Flux)/(Total Flux) ). 

The effect of the photometric errors on $\langle L_{RGB} \rangle$ is estimated by the 1$\sigma$ errors in the distribution of the $L_{RGB}$ measurements of the 1000 realizations. We combine this photometric error in quadrature with the error in distance modulus to obtain a $\pm$ 0.08 mag for And~X and 0.07 mag for And~XVII. 
However, these uncertainties do not contain the systematic errors resulting from our choice of age and alpha abundance. Thus, we recompute the total magnitude of each system using a LF for 10 Gyrs, [$\alpha$/Fe]=+0.4 and a LF for 13 Gyrs, [$\alpha$/Fe]=+0.0. Combining all uncertainties together we have resulting  
total V-band magnitudes of $M_V=-$7.36 $\pm$ 0.07 (photometric) which could inflate by $^+$ 0.29(age) $^+$ 0.63([$\alpha$/Fe]) for And~X and $M_V=-$8.61 $\pm$ 0.07 (photometric) which could inflate by $+/$ 0.02(age) $+/$ 0.28([$\alpha$/Fe]) for And~XVII.

\section{Summary}
\label{sec:summary}

Using deep broad-band observations obtained with the LBC imager on the LBT, we have derived distances, metallicities, structural parameters and total luminosities for the dwarf galaxies And~X and And~XVII. The resulting deep CMDs, presented in the left panels of Figures \ref{andx} and \ref{andxvii}, allow for the first time distance determinations to be made using the HB magnitudes of these systems. 

%We now get distances relative to M31 of 3.5$\%$. 

Based on our data, a summary of our results is as follows:

(1)   And~X is highly elliptical ($\epsilon$=0.48 $\pm$ 0.06) with its semi-major axis aligned toward Andromeda. One could wonder if tidal forces are responsible for this large elongation, however the M31-centric distance of 174 kpc makes this unlikely, but not impossible: NGC 147 has large tidal tails at  a distance of 150 kpc from M31.
In the initial discovery paper of \cite{Zucker2007}, the distance determination could only be made using the RGB tip magnitude, as deeper data were not available.  In this way a distance modulus of ($m-M$)$_0$ $\approx$ 24.12 - 24.34 was derived, yielding an absolute magnitude of $M_V$ $\approx$ $-$8.1 $\pm$ 0.5. With our revised HB distance estimate we obtain 23.98 $\pm$ 0.07, significantly closer, thus corresponding to a fainter magnitude of $M_V$ = $-$7.36$\pm$0.08, assuming [$\alpha$/Fe] =+0.4 and an age of 13 Gyr. Using this distance modulus we compute a photometric metallicity of $\langle [Fe/H]\rangle$=$-$2.2 under the assumption of [$\alpha$/Fe] =+0.4. This compares nicely with the spectroscopic results obtained  by \cite{kal09}, who found 
And~X to have [Fe/H]$_{spec}$ = $-$2.18 $\pm$ 0.15 from the co-added spectrum of 22 member red giant branch stars.
Additionally, we find a substantial error-corrected metallicity spread of 0.2$\pm$0.04 dex, strongly suggesting multiple stellar populations are present.

(2) And~XVII is a reasonably bright system with $M_V$ = $-$8.66$\pm$0.07 and a mean metallicity of $\langle [Fe/H]\rangle$=$-$2.0, under the assumption of [$\alpha$/Fe] =+0.4 and an age of 13 Gyr. 
This compares well with the properties derived from the shallow CMD in the discovery paper by \cite{Irwin2008} of [Fe/H] $\approx$ $-$1.9 and $M_{V}\approx$ $-$8.5. However we do find a closer distance modulus of 24.33 $\pm$ 0.07 compared to the ($m-M$)$_{0}$=24.5$\pm$ 0.1 found by Irwin et al. 
The revised luminosity of And~XVII is higher even though we derived a closer distance.  This is due to the fact that the luminosity is not only a function of distance, but of metallicity
and alpha enhancement as well. Here, we have revised both, contributing to 
the difference. Additionally the luminosity functions used to 
compute luminosity in \cite{Irwin2008} may have been different, however,
the author of the previous paper do not explain how they have done their computation.
After error correction, we derive the metallicity spread of And~XVII to be 0.3 $\pm$0.03 dex. This is a clear indication of a multiple stellar population system. Additionally, luminosity measurements agree with the previous determination of lower precision if systematic errors are taken into account.

 Although we confirm the bulk of the properties of And~X and And~XVII originally determined in the discovery papers by \cite{Zucker2007} and \cite{Irwin2008}, our revised HB distances put both galaxies $significantly$ closer (78 kpc for And X and  60 kpc for And XVII). Since observations of only the RGB were available at the time, \cite{Zucker2007} and \cite{Irwin2008} used the RGB tip magnitude to estimate their distances. With our new distance estimates, we show how essential HB observations are for systems with poorly sampled RGBs. Our results warn against solely using the RGB tip technique for these systems --- as missing critical stars near the tip can lead to an overestimated distance, as shown in this paper, and consequently affects both the derived metallicity and total luminosity of the system.

%%%%%%%%%%%%%%%%%%%%%%%%%%%%%%%%%%%%%% Plots
%%%%%%%%%%%%%%%%%%%%%%%%%%%%%%%%%%%

\end{document}